\pdfoutput=1

\documentclass[reprint,amsmath,amssymb,floatfix,aps,longbibliography]{revtex4-1}
\usepackage{graphics,graphicx,float}

\usepackage[pdftex,
            pdfauthor={Steven A. Frank},
            pdftitle={},
            pdfsubject={A general model of the public goods dilemma}]{hyperref} 

\def\cite{\citep}
\def\shortcite{\citep}
\def\citeA{\textcite}
\def\shortciteA{\textcite}

\newcommand{\Ga}{\alpha}
\newcommand{\Gb}{\beta}
\newcommand{\Gd}{\delta}

\DeclareMathOperator{\E}{E}
\newcommand{\dd}{{\hbox{\rm d}}}

\newcommand{\Eq}[1]{Eq.~(\ref{eq:#1})}

\newcommand{\ovr}[2]{{{#1}\over{#2}}}
\newcommand{\dovr}[2]{\ovr{\dd #1}{\dd #2}}

\begin{document}

\title{A general model of the public goods dilemma}

\author{Steven A.\ Frank}
\email[email: ]{safrank@uci.edu}
\homepage[homepage: ]{http://stevefrank.org}
\affiliation{Department of Ecology and Evolutionary Biology, University of California, Irvine, CA 92697--2525  USA}

\begin{abstract}

An individually costly act that benefits all group members is a public good.  Natural selection favors individual contribution to public goods only when some benefit to the individual offsets the cost of contribution.  Problems of sex ratio, parasite virulence, microbial metabolism, punishment of noncooperators, and nearly all aspects of sociality have been analyzed as public goods shaped by kin and group selection.  Here, I develop two general aspects of the public goods problem that have received relatively little attention.  First, variation in individual resources favors selfish individuals to vary their allocation to public goods.  Those individuals better endowed contribute their excess resources to public benefit, whereas those individuals with fewer resources contribute less to the public good. Thus, purely selfish behavior causes individuals to stratify into upper classes that contribute greatly to public benefit and social cohesion and to lower classes that contribute little to the public good.  Second, if group success absolutely requires production of the public good, then the pressure favoring production is relatively high.  By contrast, if group success depends weakly on the public good, then the pressure favoring production is relatively weak.  Stated in this way, it is obvious that the role of baseline success is important.  However, discussions of public goods problems sometimes fail to emphasize this point sufficiently.  The models here suggest simple tests for the roles of resource variation and baseline success.  Given the widespread importance of public goods, better models and tests would greatly deepen our understanding of many processes in biology and sociality\footnote{\href{http://dx.doi.org/10.1111/j.1420-9101.2010.01986.x}{doi:\ 10.1111/j.1420-9101.2010.01986.x} in \textit{J. Evol. Biol.}}.

\end{abstract}

\maketitle

\section*{Introduction}

Many biological problems turn on the dilemma of public goods.  For example, a microbe may secrete an enzyme to digest an extracellular resource into a form that can be taken up by the cell.  The cost of such secretions is borne by the individual cell that produces the enzyme.  The benefit of the secreted enzyme is, by contrast, publicly available to any neighboring cell that can take up the digested resource.  The public goods dilemma arises because those nonsecreting cells that do not produce enzymes gain the same benefit as secretors, but the nonsecretors do not pay the cost of secretion.  In direct competition, nonsecretors outcompete secretors. Enzyme secretion, as a public good, declines in frequency.

The public goods dilemma applies to any character that is directly costly to the individual and advantageous for all members of the local group.  Characters that reduce individual competitiveness or rate of resource acquisition in a way that enhances group efficiency or productivity face the dilemma.  Examples include the tradeoff between rate and yield in metabolism \shortcite{pfeiffer01cooperation}, aspects of parasite virulence \cite{lewontin70the-units}, sex ratio in local groups in which productivity depends on the number of reproductive females \cite{hamilton67extraordinary}, and contributions to repressing or punishing group members that behave selfishly \shortcite{boyd03the-evolution}. This tension between individual and group success is universal at all levels of biological organization \cite{maynard-smith95the-major}. A later section of this paper provides more detailed citations connecting public goods problems to studies in biology and the social sciences. 

In prior evolutionary models, I formulated the public goods dilemma in terms of the tragedy of the commons \cite{frank94kin-selection,frank95mutual,frank96models,frank98foundations}.   In this paper, I emphasize three key aspects of public goods problems that have not been developed in a general way. 

First, how do startup costs for production of a public good influence the amount produced. How do alternative opportunities for success in the absence of a public good influence the scaling of benefits? For example, to make a secreted enzyme, a bacterial cell must often turn on a complex pathway that requires enhanced expression of several components.  This startup cost means that production of very low levels of the public good is likely be significantly costly, whereas increasing production from low levels may not add much expense.  With regard to benefits, a public good that is essential for survival faces different pressures from a public good that only incrementally enhances success.  Such scaling issues often do not receive the attention they deserve, even though scaling may explain much of the variation in observed contributions to public goods.

Second, how do nonlinearities in costs and benefits affect production?  On the cost side, increasing production of a public good may require rising energy per unit production to drive the production at a faster rate.  On the benefit side, an excess of the secretion may cause saturation and diminishing benefits.  

Third, how does variation in resource level or vigor influence individual contribution to public goods?  In three prior studies, I found that those individuals with greater than average resource contributed more to public goods, whereas those with less than average resource reduced their contribution \cite{frank87variable,frank96policing,frank98inducible}.  Individuals rapidly stratified into an upper class that contributed greatly to public benefit and social cohesion and a lower class that contributed little to the public good.  However, those three studies were framed with regard to particular characters and particular assumptions.  Here, I analyze the problem of variable resources in a general way, to study the common processes that shape public goods dilemmas.

\section*{Public goods without individual variation}

I start with the case in which individual resources do not vary.  This case introduces the way in which natural selection shapes characters in public goods situations.  

I use the standard group-structured assumptions for biological models of social processes \cite{frank98foundations}.  The population is divided into a large number of local groups.  Interactions between individual competitiveness and group efficiency happen within local groups.  An individual's direct loss in competitiveness from contribution to public goods may be offset by its gain from the public goods contributions of its neighbors.

To start, we need an expression for individual fitness, $w$, that captures individual costliness, $c$, and group benefit, $b$, which we write as
\begin{equation}\label{eq:novary}
  w = \left[\ovr{1-c(y)}{1-c(z^*)}\right]\ovr{b(z)}{b(z^*)}.
\end{equation}
Individual production of the public good, $y$, reduces the direct individual component of fitness by the cost $c(y)$ when holding constant the group beneficial effect, $b$. We normalize the individual fitness component by $1-c(z^*)$ to get a meaningful scale for costs, where $z^*$ is the average of $y$ across all groups in the population.  The average of individual contributions to public goods within a group is $z$; the group's public goods benefit individual fitness by the group efficiency term, $b(z)$.  We normalize the benefit by the population average value, $b(z^*)$.  

We search for an evolutionarily stable strategy (ESS) $y=z=z^*$ such that, when that allocation to public goods is adopted by all members of the population, no individual can do better by deviating by a small amount from the ESS \cite{maynard-smith82evolution}.  Individual values, $y$, may be correlated with local group averages, $z$.  Such correlations may be caused by genetical kinship, by choice of group partners, or by any other process that correlates individual and group characters.  To allow for such correlations, we apply the generalized ESS methods for kin selection or correlated interactions \cite{taylor96how-to-make,frank98foundations}.  

In the general method, we search for a local maximum of the expected value of fitness for a given individual character, $\E(w|y)$, with respect to small deviations in the individual character $y$. The derivative of fitness with respect to deviations in individual character value is
\begin{equation}\label{eq:marginalrule}
  \dovr{\E(w|y)}{y} = w_y + rw_z = -C_m + rB_m,
\end{equation}
where $w_y$ and $w_z$ are the partial derivatives of fitness relative to $y$ and $z$, respectively, and $r=\dd z/\dd y$ is the slope (regression) of group character on individual character---the measure of correlation between individual and group behavior \cite{frank95mutual,taylor96how-to-make,frank98foundations}.  The terms $C_m$ and $B_m$ denote marginal costs and benefits, which allow us to see the relation to the marginal form of Hamilton's rule from the theory of social evolution \cite{taylor96how-to-make,frank98foundations}, although one must be careful when considering the meaning of Hamilton's theory of kin selection and inclusive fitness relative to the general expression of marginal costs and benefits in correlated group structures given here \cite{frank98foundations,frank06social,frank09evolutionary}.

When we apply the marginal value rule in \Eq{marginalrule} to the fitness expression in \Eq{novary}, set the derivative to zero, and solve for the ESS, we obtain
\begin{equation}\label{eq:marginalvalue}
  r\ovr{b'}{b} = \ovr{c'}{1-c},
\end{equation}
where all functions are evaluated at the ESS, $y=z=z^*$, and the primes denote derivatives.  Thus, the ESS satisfies the social rule that the marginal benefits of the group, $b'/b$, weighted by the regression of the group on the individual, $r$, must equal the marginal cost, $c'/(1-c)$.  

If we assume linear costs, $c(y) = y$, and linear benefits, $b(z)=z$, then we obtain the ESS
\begin{equation*}
  z^* = \ovr{r}{1+r}.
\end{equation*}
The value of $r$ is the regression of group value on individual value.  In groups of size $N$, the individual value is a fraction $1/N$ of the group value.  Thus, we can write $r=1/N + \hat{r}$, where $1/N$ arises from the perfect correlation of an individual to itself as a fraction $1/N$ of the group \cite{hamilton75innate,hamilton79wingless,frank83a-hierarchical,frank85hierarchical,frank86hierarchical,nunney85female-biased}, and $\hat{r}$ is the correlation between pairs of different individuals in a group \cite{frank96policing,frank98foundations}.  If the only correlation arises from an individual to itself, then $z^*=1/(N+1)$.  

\subsection*{Startup costs and fixed benefits}

To make a secreted enzyme, a bacterial cell must often turn on a complex pathway that requires enhanced expression of several components.  This startup cost means that production of very low levels of the public good is likely be significantly costly, whereas increasing production from low levels may not add much expense.  We can incorporate a startup cost by assuming $c(y)=k+y$ for $y>0$ and $c(0)=0$, where the cost of turning on the pathway is $k$.

On the benefit side, it may often be more realistic to assume some productivity in the absence of the public good. In particular, we may write the benefit as $b(z) = s + z$, so that there is a fixed productivity of $s$ in the absence of the public good.  Using these more general assumptions, we obtain the ESS contribution to public goods as
\begin{equation}\label{eq:startup}
  z^* = \ovr{r(1-k)-s}{1+r}.
\end{equation}
If $s>r(1-k)$, then the ESS is no contribution to public goods, $z^*=0$.  Thus, low correlation $(r)$ or high baseline fitness $(s)$ favors withholding of public goods, consistent with the general notion that competitive situations often disfavor contribution to public goods.

\subsection*{Nonlinearity}

On the benefit side, an excess of a public good may cause saturation and diminishing benefits.  On the cost side, increasing production of a public good requires additional raw materials and may also require rising energy per unit production to drive the production at a faster rate.  Other nonlinearities may arise.  To capture potential nonlinearities in a simple way, let $c(y) = k+y^\Gb$ and $b(z) = s+z^\Ga$. With these assumptions, we can use \Eq{marginalvalue} to obtain the ESS condition
\begin{equation*}
  z^\Gb(\Gb+r\Ga) + z^{\Gb-\Ga}s\Gb-r\Ga(1-k) = 0.
\end{equation*}
If $\Ga=\Gb$, then
\begin{equation*}
  z^* = \left[\ovr{r(1-k)-s}{1+r}\right]^{1/\Ga}.
\end{equation*}
Increasing $\Ga$ favors more allocation to public goods.  When $\Ga<1$, costs and benefits increase at a diminishing rate, and public goods allocation is less than with linear costs and benefits, suggesting that the diminishing benefits are weighted more heavily than the diminishing costs.  By contrast, when $\Ga>1$, costs and benefits increase at an accelerating rate, and public goods allocation is greater than with linear costs and benefits, suggesting that the accelerating benefits are weighted more heavily than the accelerating costs.

\section*{Public goods with individual variation}

Now suppose that individuals vary in their resource level or vigor. Each group has the same fraction of individuals with resource level class, $j$, with resources $1+\Gd_j$ and contribution to public goods $y_j$, leading to fitness as
\begin{equation}\label{eq:vary}
  w_j = \left[\ovr{1+\Gd_j-c(y_j)}{1+\Gd_j-c(z_j^*)}\right]\ovr{b(z)}{b(z^*)},
\end{equation}
where $z_j^*$ is the set of ESS values for individual contributions given the resource level of each class, $j$.  Applying the marginal social rule of \Eq{marginalrule}, we obtain for each $j$ the ESS condition
\begin{equation}\label{eq:soln}
  r_j\ovr{b'}{b} = \ovr{c_j'}{1+\Gd_j-c_j},
\end{equation}
where $r_j=\dd z/\dd y_j$, the function $b$ is evaluated at the ESS group average, $z^*$, and the function $c_j$ is evaluated at the ESS for each class, $z_j^*$.  We can see directly from this solution that, as the class correlation with the group average rises, expressed by higher $r_j$, the marginal benefit is weighted more heavily, and thus that class will be favored to change its allocation to public goods until its marginal cost rises to match the increase in the weighted marginal benefit.  In most reasonable cases, a rise in marginal cost means greater allocation to public goods. Thus, typically a rise in $r_j$ implies greater contribution to public goods. 

Given that simple role for the correlation structure set by $r_j$, let us fix $r_j=r$ for all $j$. With a fixed correlation structure, we can study how resource level affects contribution to public goods independently of the correlational structure.

If we assume linear costs, $c(y_j) = y_j$ and linear benefits $b(z) = z$, then we obtain the ESS
\begin{equation*}
  z_j^* = 1+\Gd_j - \ovr{z^*}{r},
\end{equation*}
which shows that the allocation of each class to public goods, $z_j^*$, changes directly with the class's resource level, $\Gd_j$.  If we assume that the distribution of resource levels in each group is symmetric about zero, and we assume that $z^*$ is greater than the maximum value of $\Gd_j$, then we obtain the simple expression 
\begin{equation}\label{eq:varESS}
  z_j^*-z^* = \Gd_j,
\end{equation}
where $z^*=r/(1+r)$. This simple form for the general public goods dilemma matches the solutions for particular public goods problems in earlier papers \cite{frank96policing,frank98inducible}.  

At the ESS given by \Eq{varESS}, the fitness of all classes is the same independently of their resource level. We see this by starting with a modified expression of fitness from \Eq{vary} without normalization, as $\hat{w}_j = [1+\Gd_j-c(y_j)]b(z)$.  Evaluating at the ESS, $y_j=z_j^*$ and $z=z^*$, we obtain $\hat{w}_j =(1-z^*)z^*$, which is the same for all classes and is independent of the initial resource level. 

Thus, in the public goods setting with linear costs and benefits, those with extra resources give up their entire excess for public benefit, and those with less resources withhold contribution to bring their success up to match those initially better endowed.  Each individual, in pursuing its own selfish interest, gives up its initial advantage to produce a perfectly even distribution of payoffs.

\subsection*{Startup costs and fixed benefits}

The main result in this section is that, for startup costs and fixed benefits, and with variation in resource level, we once again obtain $z_j^*-z^* = \Gd_j$, but in this case with $z^*$ from \Eq{startup}.  I give a few details.

Using $c(y)=k+y$ and $b(z) = s + z$ as before, we obtain the condition that
\begin{equation*}
  z_j^* = 1+\Gd_j-k - \ovr{s+z^*}{r}
\end{equation*}
for all $j$, with the constraint that the combination of parameters must satisfy $z_j^*\ge0$.  Assuming $z^*+\Gd_j\ge0$ for all $j$, then for a symmetric distribution of resource deviations, $\Gd_j$, centered at zero, we can use 
$z^*$ from \Eq{startup} and obtain $z_j^*-z^* = \Gd_j$.

\subsection*{Nonlinearity}

Suppose, as before, $c(y) = k+y^\Gb$ and $b(z) = s+z^\Ga$.  Then from \Eq{soln}, we obtain the ESS condition
\begin{equation*}
  \ovr{r\Ga z^{\Ga-1}}{s+z^\Ga} = \ovr{\Gb z_j^{\Gb-1}}{1+\Gd_j-k-z_j^\Gb},
\end{equation*}
for all $j$, with the constraint that all $z$ values must be nonnegative.  This equation by itself provides little insight, but is easy to evaluate numerically for particular assumptions about resource variation, startup costs, and nonlinearities.  For example, one could assume that costs rise at an accelerating rate, $\Gb>1$, because of increased energy required to drive the rate of production faster, and benefits rise at a diminishing rate, $\Ga<1$, because an increasing abundance of the beneficial public good can saturate demand or efficient usage.  

\section*{Connections to prior work}

The public goods dilemma arises in many biological, social, and economic problems.  This paper does not review the extensive theoretical literature.  However, a few pointers to prior work may be helpful. 

In biology, many problems of kin selection \cite{hamilton64the-genetical} or group selection \cite{hamilton75innate,wilson83the-group} arise from the public goods dilemma.  Since the 1990s, emphasis in the evolutionary literature on cooperation \cite{frank95mutual} and parasite virulence \cite{frank96models} have connected the public goods dilemma to Hardin's \citeyear{hardin68tragedy} slogan of ``the tragedy of the commons,'' following Leigh's \citeyear{leigh77how-does} earlier commentary. \citeA{dionisio2006tragedy} and \shortciteA{rankin07the-tragedy} recently reviewed the tragedy of the commons in biology.  \shortciteA{west07the-social} review public goods in microbes.

Nonlinearities in biology have been analyzed often.  Many of the early papers on sex allocation, tragedy of the commons, and parasite virulence included nonlinear costs and benefits.  To cite one recent example, \citeA{foster04diminishing} presented a detailed analysis of the importance of nonlinearities in understanding the tragedy of the commons.  Startup costs of production were discussed extensively for sex allocation \cite{frank87individual,west09sex-allocation}, but are not always emphasized in public goods problems. Some models do include a scaling for baseline success, such as West \textit{et al.}'s \citeyear{west2002sanctions} study of nitrogen fixation by rhizobia.

Variable resources have been developed most extensively for problems of sex allocation \cite{werren80sex-ratio,yamaguchi85sex-ratios,frank87individual,frank87variable,west09sex-allocation}, but that work did not make a direct connection to the general public goods dilemma.  My work on variable resources in repression of competition \cite{frank96policing} and herd immunity \cite{frank98inducible} was within a public goods context, but I did not emphasize the generality of the solution and the connection to the earlier analyses of sex allocation.  There must be several other studies of variable resources related to public goods problems. However, there is a tendency in the biological literature to ignore variability even though it is both widespread and important.  

Public goods and the tragedy of the commons are discussed widely in the social sciences.  \citeA{ostrom77collective,ostrom90governing,ostrom99coping} has contributed extensively to conceptual analysis and empirical application.  Economic theory has a highly developed literature on a variety of related topics, some of which are described as public goods problems \cite{cornes09voluntary}.  Those problems are typically framed in somewhat different ways from the simple biologically motivated models in this paper, but some economic work does turns on individual versus group tension.  Spatial population structure is one key difference between the biological and social science models: often, the social science models assume a single population, and so do not include a correlation between individual and group behavior.  Without correlation, $r=0$, it is more difficult to achieve individual contribution to public goods. Boyd and Richerson's \citeyear{boyd02group} work on cultural evolution in group structured populations may provide an interesting connection between problems in the biological and social sciences.  

\section*{Discussion}

I emphasize two conclusions.  First, variation in individual resources favors selfish individuals to vary their allocation to public goods.  Those individuals better endowed contribute their excess resources to public benefit, whereas those individuals with fewer resources contribute less to the public good.  In the simplest case, all individuals end up with the same fitness in spite of initial variation in endowment.  Thus, purely selfish behavior causes individuals to stratify into upper classes that contribute greatly to public benefit and social cohesion and to lower classes that contribute little to the public good.

The role of variation may be tested experimentally in microbes. One could manipulate resources by creating variants with and without the ability to use different sources of energy.  By controlling the abundance of each energy source, one could create different classes of individuals that have access to different levels of resource.  The theory here predicts that the microbes would evolve a stratified pattern of contribution to public goods.  Similar behavioral tests in primates may also be possible. However, the complexity of behavioral strategies may make it difficult to sort among alternative hypotheses for changes in behavior in response to manipulated or natural variation in resource level.

The second conclusion concerns the group's dependence on the public good.  If group success absolutely requires production of the public good, then the pressure favoring production is relatively high.  By contrast, if group success depends weakly on the public good, then the pressure favoring production is relatively weak.  I expressed these dependencies by writing the benefit term as $b(z) = s+z$, where $s$ sets the level of group benefit in the absence of the public good.  Once stated in this way, it is obvious that the role of baseline success is important.  However, discussions of public goods problems sometimes fail to emphasize this point sufficiently.

The role of baseline success may be tested by manipulating the group's dependence on production of a public good.  In experimental evolution studies of microbes, one could measure the response of public goods productivity to changes in dependence measured by $s$.  For example, \shortciteA{kummerli09phenotypic} experimentally manipulated $s$ by altering the amount of iron available to bacteria and studying the evolutionary response of secreted iron-scavenging public goods molecules.  Their results support the prediction that as baseline fitness rises with increasing experimentally provided iron, bacteria reduce their contribution to the costly public good.  Observational studies of microbes may be able to compare natural settings in which dependence on a public good varies.  Experiments or observational studies in primates or other animals would also be possible, but once again it may be difficult to separate between alternative hypotheses with regard to behavioral changes.  

\section*{Acknowledgments}

My research is supported by National Science Foundation grant EF-0822399, National Institute of General Medical Sciences MIDAS Program grant U01-GM-76499, and a grant from the James S.~McDonnell Foundation.  

\vfill\eject

\bibliography{microbe}

\vfill\eject

%
%
%
%
%

\end{document}